\documentclass[twocolumn]{jpsj3}
\usepackage{txfonts}
\usepackage{color,graphicx}

\title{Quantum Theory of Rare-Earth Magnets}

\author{
Takashi Miyake$^{1,3}$\thanks{t-miyake@aist.go.jp} 
and Hisazumi Akai$^{2,3}$
}
\inst{
$^1$CD-FMat, National Institute of Advanced Industrial Science
and Technology, Tsukuba 305-8568, Japan \\
$^2$The Institute for Solid State Physics, The University of
Tokyo, 5-1-5 Kashiwano-ha, Chiba 277-8581, Japan \\
$^3$ESICMM, National Institute for Materials Science, Tsukuba 305-0047, Japan} 

\abst{
Strong permanent magnets mainly consist of rare earths ($R$) and transition metals ($T$). The main phase of the neodymium magnet, which is the strongest magnet, is Nd$_2$Fe$_{14}$B. Sm$_{2}$Fe$_{17}$N$_{3}$ is another magnet compound having excellent magnetic properties comparable to those of Nd$_{2}$Fe$_{14}$B. Their large saturation magnetization, strong magnetocrystalline anisotropy, and high Curie temperature originate from the interaction between the $T$-3d electrons and $R$-4f electrons. This article discusses the magnetism of rare-earth magnet compounds. The basic theory and first-principles calculation approaches for quantitative description of the magnetic properties are presented, together with applications to typical compounds such as Nd$_2$Fe$_{14}$B, Sm$_{2}$Fe$_{17}$N$_{3}$, and the recently synthesized NdFe$_{12}$N.
}

\begin{document}
\maketitle

\section{Introduction} 
\label{sec:introduction}
Modern permanent magnets are the consequences of the fine combination of various magnetic and nonmagnetic materials, as well as micro-, macro-, and metallographic structures \cite{Coey2010}. Thus, quantum theory tells only part of the story of rare-earth magnets. Nevertheless, since magnetism is one of the most prominent manifestations of the quantal nature of electrons \cite{VanVleck1932}, quantum theory must be a key player in studying permanent magnet materials.  In this review, we will concentrate mostly on the electronic and magnetic properties of single crystals of rare-earth magnet materials, discussing some selected topics that may be essential in terms of developing permanent magnets.

Rare earths have particular importance in modern permanent magnets. The reason why rare-earth elements are so important in magnets is that one of the necessary conditions for a ferromagnet to be a permanent magnet material is magnetic anisotropy. The magnetic anisotropy originates from either crystalline or shape anisotropy; the latter can never be strong enough for modern magnets. The former is the result of spin-orbit coupling (SOC) which eventually sticks spins to a crystal structure. The strength of single-electron SOC of 4f  electrons of rare earths is $\sim 0.5$ eV. For Fe-3d electrons, it is one order of magnitude smaller. Although rare-earth magnets contain only a small amount of rare earths, e.g., less than 1/7 of the whole in the case of Nd$_2$Fe$_{14}$B magnets, adding them enhances the magnetic anisotropy at the working temperature by $\sim 50$\% (naturally much more at low temperatures), which is already a huge increase from the technological point of view.

Unfortunately, as is widely recognized, there is no established way to treat the electronic and magnetic properties of 4f electron systems from first principles. This makes the theoretical treatment of rare-earth magnets rather difficult. At best, what we can do now is to compromise and to add some kinds of ad hoc treatments, under several assumptions,  each of them being not based on an approximation of the  same level, on top of the standard first-principles theory. In the subsequent sections, we will review the recent development of quantum mechanical approaches to the problem of rare-earth magnets, which mostly follow such types of incomplete approaches. 

In Sect.~\ref{sec:basics}, we briefly review typical rare-earth magnets and the basic idea of their magnetism. 
The framework of first-principles approaches that are used to describe rare-earth magnets is explained in Sect.~\ref{sec:method}. Some examples of the approaches are also given. In particular, the magnetic anisotropy of Sm$_2$Fe$_{17}$N$_x$ is discussed in detail. The finite-temperature properties of rare-earth magnets are discussed in Sect.~\ref{sec:finiteT} together with some methodological aspects. Several different but complementary methods are explained with some recent results. Section \ref{sec:RFe12} deals with NdFe$_{12}$N and related compounds. NdFe$_{12}$N has been synthesized recently, and turned out to have excellent intrinsic magnetic properties surpassing those of Nd$_{2}$Fe$_{14}$B. 
Section \ref{sec:summary} summarizes the review.

\section{Rare-Earth Magnet Compounds}
\label{sec:basics}
\begin{figure}[htp]
\includegraphics[width=9cm]{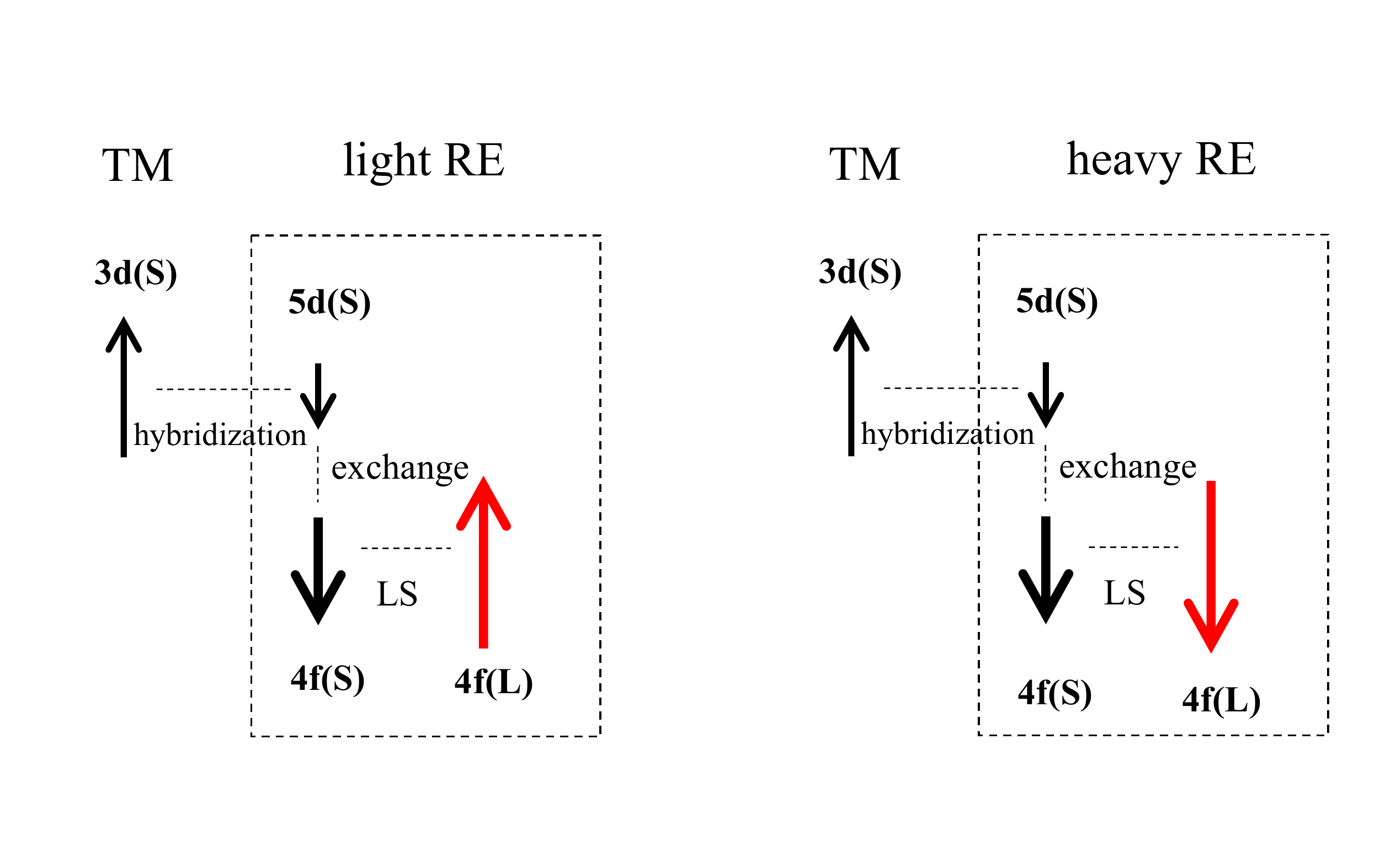}
\caption{(Color online) 
Interaction between 3d electrons of transition metals and rare-earth 4f electrons in rare-earth magnet compounds. 
}
\label{fig:3d4f}
\end{figure}

Rare-earth magnet compounds \cite{buschow1977,li1991} are mainly composed of  transition-metal ($T$) and rare-earth ($R$) elements. The majority component is 3d transition metals, which are essential for a large saturation magnetization and high Curie temperature, while $R$ elements are responsible for strong magnetocrystalline anisotropy. Figure~\ref{fig:3d4f} shows the electronic states schematically. As the electron configuration of $R$-4f electrons follows Hund's rule, the orbital magnetic moment appears in the presence of SOC. The resultant electron distribution slightly deviates from a spherical shape. The nonspherical component is subjected to a crystal electric field produced by other electrons and ions, which determines the direction of the 4f orbital moment. Once the direction is fixed, the direction of the spin moment is also fixed by the LS coupling. The 4f electrons are coupled to 5d electrons by the intraatomic exchange interaction, consequently, their spin moments are aligned parallel to each other. Since the 5d orbitals are spatially extended, they hybridize with the $T$-3d orbitals antiferromagnetically. Therefore, the $T$-3d spin is antiparallel to the $R$-4f spin. 

The above consideration indicates that the strength of the crystal field at the rare-earth site is a good measure of the  magnetocrystalline anisotropy of a rare-earth magnet compound. In crystal-field theory \cite{stevens1952,huttings1964}, the magnetic anisotropy constant $K_1$, defined by 
\begin{equation}
E(\theta) \simeq K_1 \sin^2\theta \;,
\label{eq:MCA}
\end{equation}
can be expressed as 

\begin{equation}
K_1 = -3 J (J-1) \alpha_J \langle r^2 \rangle A_2^0 n_R
\;,
\label{eq:K1}
\end{equation}
where $\theta$ is the angle between the easy axis and the magnetization, 
$J$ is the total angular momentum, 
$\alpha_J$ is the first Stevens factor, 
$\langle r^2 \rangle$ is the spatial extent of the 4f orbital, 
$A_2^0$ is the second-order crystal-field parameter, 
and $n_R$ is the rare-earth concentration. Here, $J$ and $\alpha_J$ are constants depending on the $R$ ion, while $A_2^0$ reflects the electronic states of the compound. 

Table \ref{tab:Rion} shows the ground-state properties of trivalent rare-earth ions. The orbital angular momentum (${\bf L}$) is parallel (antiparallel) to the spin angular momentum (${\bf S}$) when the 4f shell is more (less) than half-filled as shown in Fig.~\ref{fig:3d4f}. Hence, the total angular momentum ${\bf J} = {\bf L}+{\bf S}$ has a larger magnitude in heavy rare-earth elements than in light rare-earth elements. This is the reason why heavy rare-earth elements, e.g., Dy,  enhance the magnetocrystalline anisotropy of a rare-earth magnet. Meanwhile, the magnetic moment is suppressed as the heavy rare-earth concentration increases since the 4f magnetic moment partially cancels the magnetic moment of the $T$-3d electrons. Therefore, heavy rare-earth elements generally suppress the performance of a permanent magnet, although they improve coercivity  through enhancement of the magnetocrystalline anisotropy. The first Stevens factor $\alpha_J$ is also shown in Table \ref{tab:Rion}. A positive (negative) $\alpha_J$ means that the electron distribution of the 4f electrons is elongated (compressed) in the direction of the orbital moment, and the magnitude of $\alpha_J$ gives the degree of asphericity. We see that the sign of $\alpha_J$ of Nd$^{3+}$ is opposite to that of Sm$^{3+}$. Assuming that the crystal-field parameter $A_2^0$ is insensitive to the $R$ ion, as a rule of thumb, a Nd-based compound shows uniaxial (basal-plane) anisotropy when the corresponding Sm-based compound has basal-plane (uniaxial) anisotropy. 

The situation is more complicated in some cases. For example, cerium has a mixed-valence state. There are no f electrons in most Ce-based magnet compounds, hence, a simple argument based on Eq. (\ref{eq:K1}) does not hold. A Ce compound will possess strong magnetocrystalline anisotropy if we can make Ce trivalent because the magnitude of the Stevens factor is large. Samarium is also a difficult element to treat theoretically. Because the energy splitting between the $J$ multiplets is small, excited $J$ states would affect the finite-temperature magnetism of Sm systems. The hybridization effect between 4f and other orbitals is also to be considered, which will be discussed in detail in the following section by taking Sm$_{2}$Fe$_{17}$N$_{3}$ as an example. Contributions from transition-metal sublattices to magnetocrystalline anisotropy are another factor, which will be discussed in Sect.~\ref{sec:finiteT}.

\begin{table}
\begin{center}
\begin{tabular}{ccccccc}
\hline
\hline
$R$ ion & f$^n$ & $L$ & $S$ & $J$ & $g_J$ & $\alpha_J$ \\
\hline
Ce$^{3+}$ & f$^1$ & 3 & 1/2 & 5/2 & 6/7 & $-2/(5 \cdot 7)$ \\
Pr$^{3+}$ & f$^2$ & 5 & 1 & 4 & 4/5 & $-2^2 \cdot 13 / (3^2 \cdot 5^2 \cdot 11)$ \\
Nd$^{3+}$  & f$^3$ & 6 & 3/2 & 9/2 &8/11 &  $-7 / (3^2 \cdot 11^2)$ \\
Pm$^{3+}$  & f$^4$ & 6 & 2 & 4 & 3/5 & $2 \cdot 7 / (3 \cdot 5 \cdot 11^2)$ \\
Sm$^{3+}$  & f$^5$ & 5 & 5/2 & 5/2 & 2/7 & $13 / (3^2 \cdot 5 \cdot 7)$ \\
Eu$^{3+}$  & f$^6$ & 3 & 3 & 0 & 0 &  0\\
Gd$^{3+}$  & f$^7$ & 0 & 7/2 & 7/2 & 2 & 0 \\
Tb$^{3+}$  & f$^8$ & 3 & 3 & 6 & 3/2 & $-1 / (3^2 \cdot 11)$ \\
Dy$^{3+}$  & f$^9$ & 5 & 5/2 & 15/2 & 4/3 & $-2 / (3^2 \cdot 5 \cdot 7)$ \\
Ho$^{3+}$  & f$^{10}$ & 6 & 2 & 8 & 5/4 & $ -1 / (2 \cdot 3^2 \cdot 5^2)$ \\
Er$^{3+}$  & f$^{11}$ & 6 & 3/2 & 15/2 & 6/5 & $ 2^2 / (3^2 \cdot 5^2 \cdot 7)$ \\
Tm$^{3+}$  & f$^{12}$ & 5 & 1 & 6 & 7/6 & $ 1 / (3^2 \cdot 11)$ \\
Yb$^{3+}$  & f$^{13}$ & 3 & 1/2 & 7/2 & 8/7 & $ 2 / (3^2 \cdot 7)$ \\
\hline
\hline
\end{tabular}
\end{center}
\caption{Number of f electrons, orbital momentum $L$, spin momentum $S$, total angular momentum $J$, 
the Land\'{e} g factor and the first Stevens factor $\alpha_J$ of trivalent rare-earth ions.}
\label{tab:Rion}
\end{table}

\begin{figure*}[htp]
\includegraphics[width=18cm]{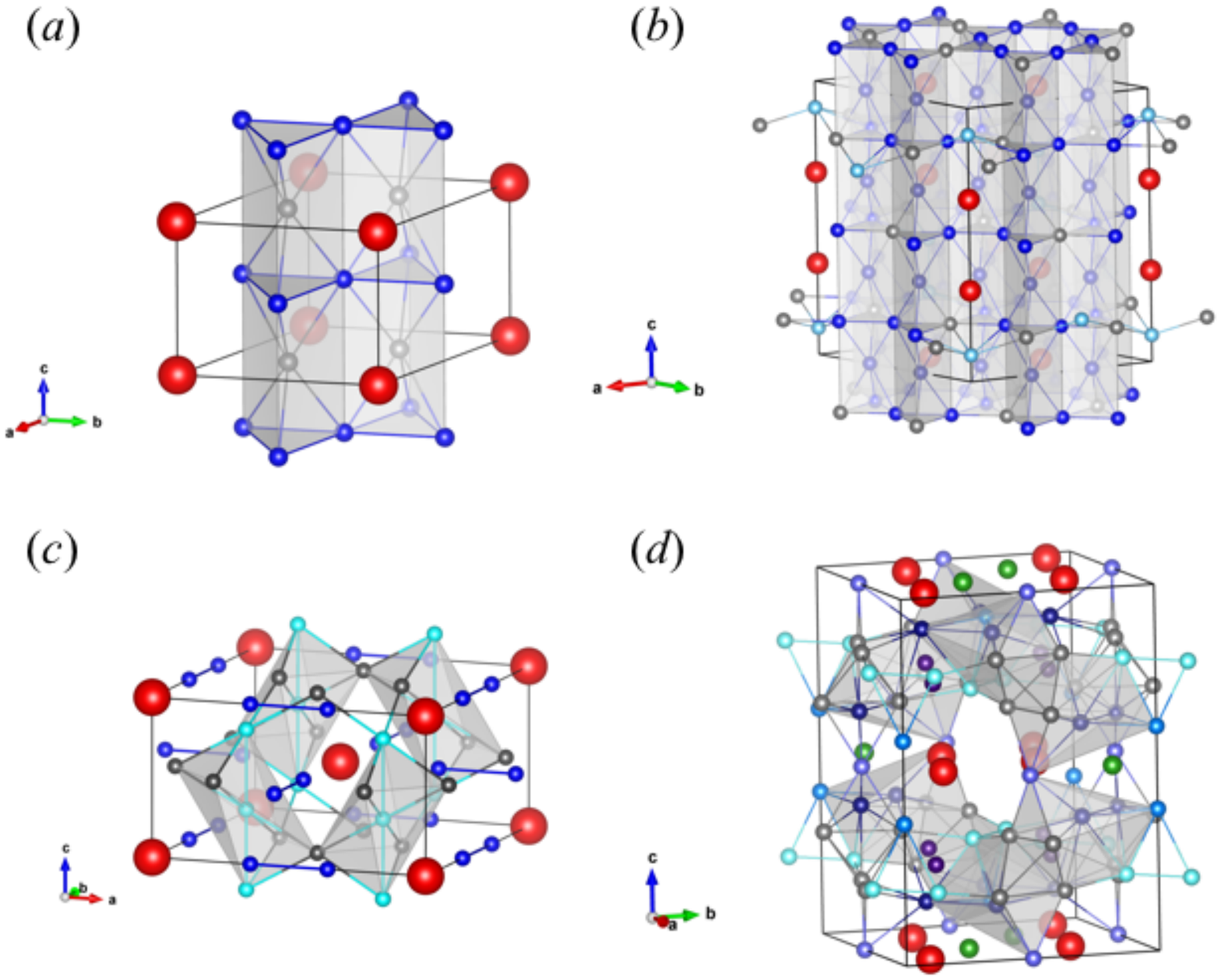}
\caption{(Color online) 
Crystal structures of rare-earth magnet compounds.
(a) CaCu$_5$ structure, 
(b) Th$_2$Zn$_{17}$ structure, 
(c) ThMn$_{12}$ structure, and 
(d) Nd$_2$Fe$_{14}$B structure. 
The red balls represent rare-earth atoms. The green balls in (d) are boron. 
Other balls represent transition metals.
}
\label{fig:structure}
\end{figure*}

The rare-earth magnet compounds are classified into several families depending on their chemical composition. The simplest one is $RT_5$ having the CaCu$_5$ structure [Fig.~\ref{fig:structure}(a)]. It has a hexagonal unit cell containing one formula unit. There are two $T$ sites, 2$c$ and 3$g$. The 2$c$ sites form a honeycomb lattice, and $R$ is located at the center of a hexagon. The 3$g$ sites form a kagom\'{e} lattice. Because of this characteristic crystal structure, flat bands exist in the electronic band dispersion \cite{ochi2015}. The first-generation rare-earth magnets YCo$_5$ \cite{hoffer1966} and SmCo$_5$ \cite{strnat1967} belong to this family. 

By replacing $n$ out of $m$ rare-earth sites with a pair of transition-metal sites (``dumbbell"), $R_{m-n}T_{5m+2n}$ is obtained. There are two structures for $(m,n)$=(3,1). One is the rhombohedral Th$_2$Zn$_{17}$ structure shown in Fig.~\ref{fig:structure}(b). The other is the hexagonal Th$_{2}$Ni$_{17}$ structure. In the former case, the dumbbell $T$ sites are arranged in a sequence of $ABCABC$ along the $c$ direction, while the stacking sequence is $ABABAB$ in the latter structure. Sm$_2$Co$_{17}$ with the Th$_2$Zn$_{17}$ structure belongs to this $R_{2}T_{17}$ family. Since it contains a higher Co content than SmCo$_5$, the saturation magnetization is larger.  So is $(BH)_{\rm max}$, although its magnetocrystalline anisotropy is weaker. While Sm$_{2}$Co$_{17}$ is a strong magnet compound, Sm$_{2}$Fe$_{17}$ shows basal-plane anisotropy. However, the anisotropy is changed by adding nitrogen. Interstitial nitrogenation induces strong uniaxial magnetic anisotropy as well as an increase in the magnetization and Curie temperature \cite{coey1990,iriyama1992}. 

If half of the $R$ sites in $RT_5$ are substituted with dumbbell $T$ pairs, $RT_{12}$ is obtained [$(m,n)$=(2,1)]. The crystal structure is ThMn$_{12}$-type with the body-centered tetragonal structure [Fig.~\ref{fig:structure}(c)]. This family has been studied intensively in recent years. Notably, it was reported that NdFe$_{12}$N has a larger saturation magnetization and anisotropy field than Nd$_2$Fe$_{14}$B \cite{hirayama2015}. We will discuss this family in Sect.~\ref{sec:RFe12}. 

Both Sm$_{2}$Fe$_{17}$N$_{3}$ \cite{iriyama1992,coey1990} and NdFe$_{12}$N contain nitrogen. In fact, light elements  vary the magnetic properties of rare-earth transition-metal intermetallics. Historically, boron provided a breakthrough before the development of nitrogenated systems. By adding boron to Nd$_{2}$Fe$_{17}$, Sagawa et al. invented a sintered neodymium  magnet whose main phase was Nd$_{2}$Fe$_{14}$B \cite{sagawa1984}. Croat et al. independently developed a melt-spun neodymium magnet \cite{croat1984}. The neodymium magnet has been the strongest magnet in the last three decades. Sagawa's intention was to raise the Curie temperature of Nd$_{2}$Fe$_{17}$ by inserting boron. He thought that boron would increase the Fe-Fe distance, which may lead to an increase in the Curie temperature. Indeed, the Curie temperature was raised, but the microscopic mechanism was different from what Sagawa had expected. The chemical formula was not Nd$_2$Fe$_{17}$B$_{x}$ but Nd$_{2}$Fe$_{14}$B, and the crystal structure was a complicated tetragonal one containing four formula units in a unit cell [Fig.~\ref{fig:structure}(d)]. 

\section{First-Principles Calculation}
\label{sec:method}
First-principles electronic structure calculation commonly means calculation based on the local density approximation (LDA) or its slight extension, the generalized gradient approximation (GGA), within the framework of density functional theory (DFT) \cite{hohenberg1964,kohn1965}.  A standard theoretical investigation of the properties of condensed matter starts from a first-principles electronic structure calculation as a first step. This is particularly true for magnetic materials since ground-state magnetic properties are well treated by first-principles calculations in most cases. However, for rare-earth magnets, such an approach fails in many cases. This is explained in the following. 

When applied to the element magnets Fe, Co, Ni, both the LDA and GGA produce mostly reasonable results. An exceptional case is the LDA applied to Fe: the LDA predicts incorrect crystal and magnetic structures for the ground state of Fe. The failure of the LDA for Fe, however, simply originates from the fact that the LDA has a tendency of overbinding the atoms, predicting too small an equilibrium lattice constant of Fe. In this sense, it is not so fatal as might be supposed. The applicability of the LDA/GGA is also similar for various magnetic intermetallic compounds and compounds such as transition-metal chalcogenides, pnictides, and halides. In the vicinity of the border separating the high-spin state from the low-spin (or nonmagnetic) state, and also near the region where the metal-insulator transition occurs, these approaches often fail to predict the correct ground states\cite{Martin2016}. However, we may say that the LDA/GGA calculation correctly describes the overall behavior of these materials in general.

The above is not true for rare-earth magnets: the f-states of rare-earth elements cannot be properly treated in the framework of the LDA/GGA. For example, let us consider SmCo$_5$, which is a prototype rare-earth permanent magnet compound. Sm has five f-electrons in its trivalent state as is usually the case in a crystal. The corresponding atomic LS multiplet of the lowest energy is $^6H_{5/2}$. Even in a crystal, such an atomic configuration of Sm 4f electrons is well preserved since the hybridization of the 4f state with the f-symmetry states composed of s, p, and d states of neighboring sites is  fairly small. In this situation, together with the fact that the f-states must be partially occupied, what would be expected is that all the 4f states are pinned at the Fermi level. This implies that the energy required for valence fluctuation to take place is quite small and the  LS multiplet loses its meaning. In reality, for these narrow states, the effects of the electron-electron interaction are so strong that the electronic states are not any more extended. The electronic states split into occupied and unoccupied states, the former being pushed down rather deep inside the Fermi sea and the latter being pushed up above the Fermi sea. This situation can hardly be reproduced by the LDA/GGA, where all the Kohn--Sham orbitals are generated with a single common effective potential. Since there are no schemes that improve this situation in a fundamental way, even state-of-the-art calculations have to solve the problem in an adhoc way.

There are several easy fixes. The first one is the ``open-core" approach, where the f states are dealt with as open-shell core states. In all electron approaches such as the full-potential linearized augmented plane wave (FLAPW) method and the Korringa--Kohn--Rostoker (KKR) Green's function method, these core states with positive energy eigenvalues have to be calculated explicitly with a rather artificial  boundary condition, e.g., a zero or zero-derivative boundary condition on radial wave functions. Suitable care has to be taken so as not to include these core states within the valence f-states. In the KKR, this can be done by removing the resonances, which  correspond to the virtual-bound f states from the atomic t-matrix. Another possible way to obtain open cores is to simply shift the potential for the f-states downward so that the energy eigenvalue could be negative even under a natural boundary condition. In pseudopotential codes, open-core treatment is easily done by including the f-states as open-shell core states when constructing pseudopotentials.

The second approach is to apply self-interaction corrections (SIC) to the f-states. This scheme obviously is beyond the scope of density functional theory in a strict sense. Nevertheless, it can be a reasonable approach if the targeted states actually localize. For such localized states, at least the self-exchange energy can be calculated exactly and hence may give a better description of the exchange-correlation energy. The SIC causes orbital splitting as naturally expected, and remedy the shortcoming of the LDA/GGA.

The third one is the so-called LDA+U method \cite{anisimov1997}, which is nothing but a local Hartree--Fock approximation. If the f-states are known to be localized, which is the same situation as needed for SIC to be applicable, the local Hartree--Fock approximation might not be a bad approximation. The method, however, is not one that takes account of the ``strong correlation": it merely introduces the effects of a strong electron interaction by hand. 

\begin{figure}[htp]
\includegraphics[width=6cm,angle=90]{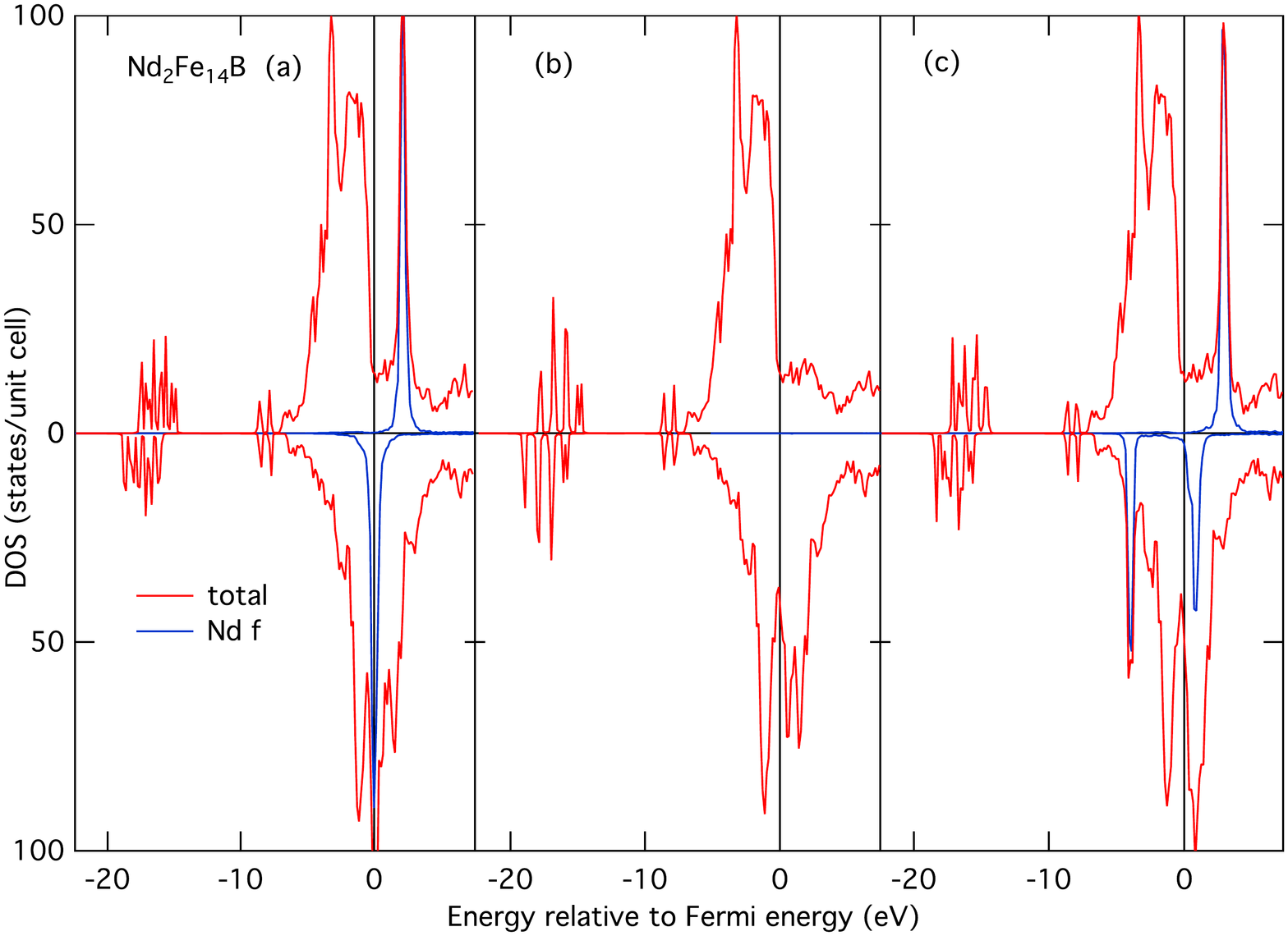}
\caption{(Color online) 
Density of states of Nd$_2$Fe$_{14}$B  calculated using (a) GGA (PBE), (b) open-core scheme, and  (c) SIC. The spin-down core states of Nd in (b) appear slightly above the Fermi level. The Nd local f densities of states are indicated by blue lines. The states around $-18$ eV are the Nd semicore p-states. 
}
\label{fig:ndfeb_dos}
\end{figure}

As an example, Fig. \ref{fig:ndfeb_dos} shows the calculated density of states of Nd$_2$Fe$_{14}$B using the GGA, the open-core, and the SIC scheme, respectively. All the calculations are performed using KKR codes, where the open-core and SIC schemes are implemented with the GGA (PBE) exchange-correlation energy \cite{perdew1996}. The spin-orbit coupling is included on top of the scalar relativistic approximation.

The resulting magnetic properties are compared in Table \ref{tab:moment}. Note that the calculated spin and orbital magnetic moments are the values projected to $(L, M_L, S, M_S)$ states. Therefore, the magnetic moment that can be obtained from the total angular momentum $J$ would be slightly larger in the present case. For example,  $M_{\rm s}$ obtained for the open-core calculation is 1.90 T instead of 1.87 T if the spin-orbitals be used. 

\begin{table}
\vskip 16pt 
\begin{center}
\begin{tabular}{cccccc}
\hline
\hline
Scheme & $M_{\rm spin} (\mu_{\rm B})$&$M_{\rm orb} (\mu_{\rm B})$&$M_{\rm s}$(T)&$T_{\rm C}$ (K)&$E_{\rm total}$ (Ry) \\
\hline
GGA & 100.9 & 31.3 & 1.63 &1184 & -296731.6327\\
open-core & 100.8 & 50.7 & 1.87 & 1157 & -296730.3632\\
SIC&100.1 & 49.6 & 1.85 & 1160 & -296731.2779\\
\hline
\hline
\end{tabular}
\end{center}
\caption{Spin magnetic moment ($M_{\rm spin}$), orbital magnetic moment ($M_{\rm orb}$), saturation magnetization ($M_{\rm s}$), Curie temperature ($T_{\rm C}$), and total energy ($E_{\rm total}$) of Nd$_2$Fe$_{14}$B calculated by three different treatments of Nd f-states: treating them as valence states (GGA),  as partially occupied core states  (open-core), and with self-interaction correction (SIC).}
\label{tab:moment}
\end{table}

The magnetic properties do not depend strongly on the schemes except for the orbital magnetic moment  $M_{\rm orb}$. This indicates that the f states of Nd do not contribute significantly in determining the magnetic properties as  a whole. However, as is implied by the considerable difference in $M_{\rm orb}$, the magnetic anisotropy, which is not calculated here, could be affected by the scheme of treatment. It is also pointed out that the equilibrium lattice constants depend on the treatment of the f states (in the above calculation the lattice constants are fixed to the experimental values). In general, the LDA gives the smallest equilibrium volume. The volume becomes larger for the GGA, SIC, and open-core treatments in this order. The volume change naturally affects the magnetic properties considerably. 

Values of Curie temperature $T_{\rm C}$ are estimated by the mean field approximation assuming a Heisenberg model; the exchange coupling constants $J_{ij}$ are calculated using the scheme obtained by Oguchi et al. \cite{oguchi1983} and by Liechtenstein et al. \cite{liechtenstein1987} It should be noticed that $T_{\rm C}$ depends on details of the calculations as well as the choice of the exchange-correlation energy. For example, if the LDA were used, $T_{\rm C}$ would be 1049 K instead of the GGA(PBE) value of 1184 K.

One of the strategies in developing permanent magnet materials is improving their performance by forming alloys. For example, a common way to improve the high temperature performance of Nd$_2$Fe$_{14}$B is to introduce some Dy that substitutes for Nd. The calculation of the electronic structure of such substitutional alloys can be conveniently performed in the framework of the coherent potential approximation (CPA). Such types of calculations are also possible for {\it substitutional alloys between vacancies and atoms}, Vc$_{1-x}$A$_x$, where Vc indicates vacancy. An example is Sm$_2$Fe$_{17}$N$_x$  ($0 \le x \le 3$), where N randomly occupies one of the three vacant interstitial sites adjacent to Sm. 

In the following, we review the results of recent calculations on Sm$_2$Fe$_{17}$N$_{x}$ \cite{ogura2015}. Sm$_2$Fe$_{17}$N$_3$ with the Th$_2$Zn$_{17}$ structure shows a much larger magnetic anisotropy and a higher Curie temperature than Nd$_2$Fe$_{14}$B, although its saturation magnetization is slightly smaller than that of the latter. Experimentally, adding N to Sm$_2$Fe$_{17}$ increases the saturation magnetization by 12\% and the Curie temperature by 93\%, and changes its magnetic anisotropy from in-plane to uniaxial, thus making it suitable for a permanent magnet material \cite{coey1990}. 
Unfortunately, Sm$_2$Fe$_{17}$N$_3$  decomposes at high temperatures\cite{Katter1992}, which prevents us from producing sintered magnets.  For this reason, it has never replaced Nd$_2$Fe$_{14}$B. However, studying Sm$_2$Fe$_{17}$N$_3$ will provide us with some hints that might be useful when seeking new high-performance permanent magnet materials.

The electronic structure was calculated by using the KKR Green's function method with the LDA (MJW parametrization) of  density functional theory. The relativistic effects are taken into account within the scalar relativistic approximation.
The SOC (only the spin diagonal terms) is included. The SIC scheme for the Sm-f states is exploited.
The nonstoichiometric content of N is treated as mentioned above using the CPA, i.e., the 9$e$ site in the Th$_{2}$Zn$_{17}$ structure is randomly occupied by N or a vacancy with the probability corresponding to the content of N.
Three types of different sets of lattice parameters are used: structure A has the experimental parameters of Sm$_2$Fe$_{17}$N$_3$\cite{inami2014}, structure B has those of Sm$_2$Fe$_{17}$\cite{teresiak2001}, and structure C has the same volume as structure A and the same atomic positions as structure B.
The atomic sphere approximation (ASA) is employed and the ratio of the radii among them is taken to be 1:1:0.5 for Sm:Fe:N. Note that this ratio sometimes affects the results considerably. In the following calculation the ratio was not in particular adjusted.
The maximum angular momentum of the atomic scattering t-matrix of KKR is 3 for Sm and 2 for the others. Higher angular momenta are taken into account as non-scattering states that contribute in determining the Fermi level.

\begin{figure}[htp]
\begin{center}
\includegraphics[width=5cm]{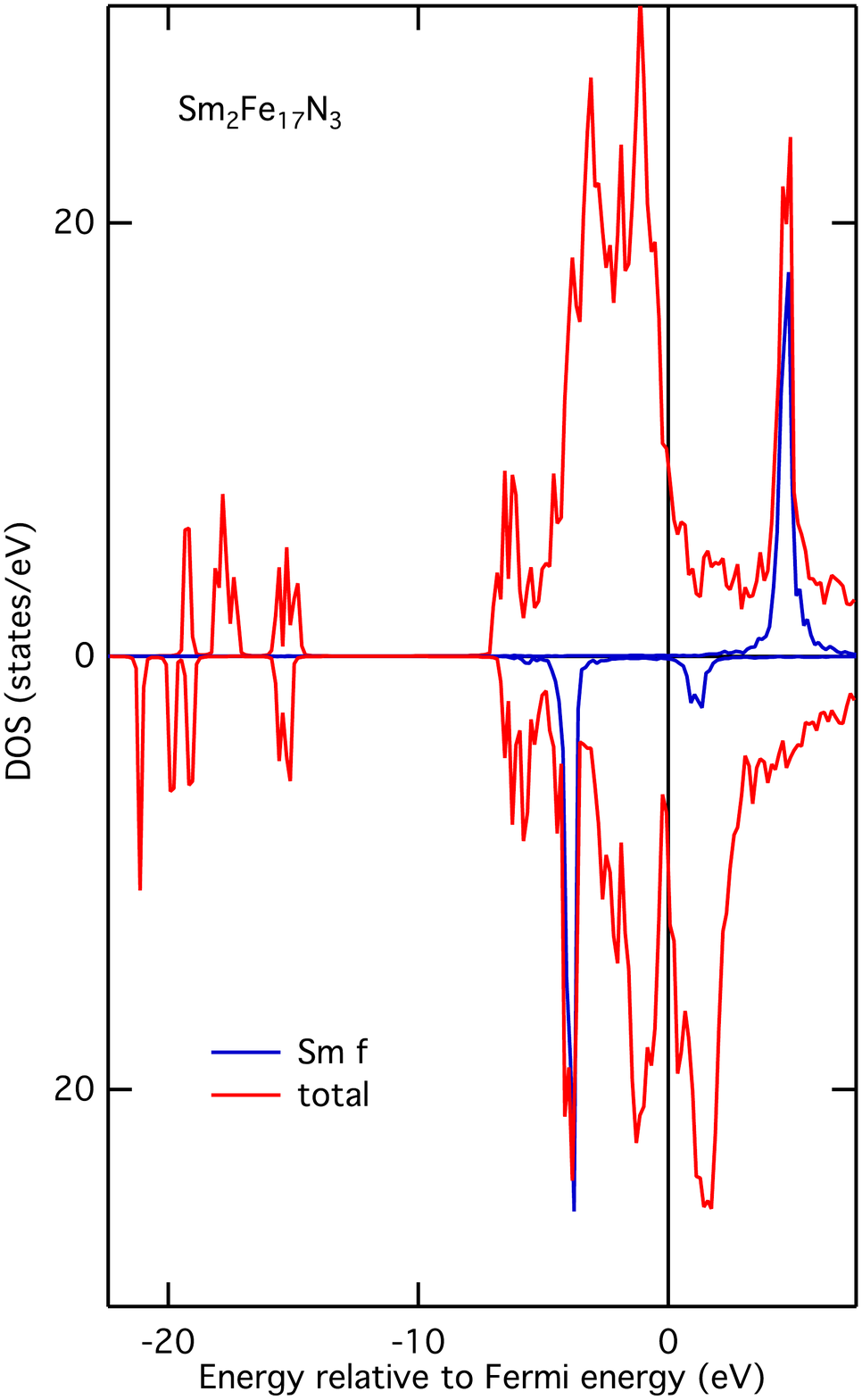}
\caption{(Color online) 
Density of states of Sm$_2$Fe$_{17}$N$_3$ calculated using LDA with SIC. Sm local f densities of states are indicated by blue lines. The states around $-18$ eV originate from the Sm semicore 5p states, those around -15 eV originate from the N semicore 2s states.
}
\label{fig:smfen_dos}
\end{center}
\end{figure}

Figure~\ref{fig:smfen_dos}  shows the density of states of Sm$_2$Fe$_{17}$N$_3$ with structure A.
Sm-f spin-down states split into two parts. The occupied f states further split, showing a finer structure. The total number of electrons in the occupied Sm-f states is 5.76 when calculated using the above mentioned ratio of the ASA radii, and thus the configuration is more or less Sm$^{2+}$. This result is rather definite and also consistent with the results obtained by LDA calculations, although the occupied f states are located at an energetically much deeper position than those in the LDA result. The fact that Sm is likely to be divalent contradicts the usual assumption that Sm is more or less trivalent. However,  we have to be particularly cautious about the {\it valency} for metallic systems such as Sm$_2$Fe$_{17}$N$_3$. Firstly, although f states are fairly localized, they still have positive energies and extend to the interstitial region. Thus, the number of  f electrons strongly depends on the volume assigned to the Sm atom, while the volume itself is to some extent arbitrary. Second, for such systems, we do not know how to define the valency that corresponds to the concept of {\it chemical valency} in the chemistry sense. Therefore, the best we can do is to compare the predicted and observed spectroscopic data that may reflect the electron configuration, without asking about the valency.

\begin{figure}[htp]
\begin{center}
\includegraphics[width=7cm]{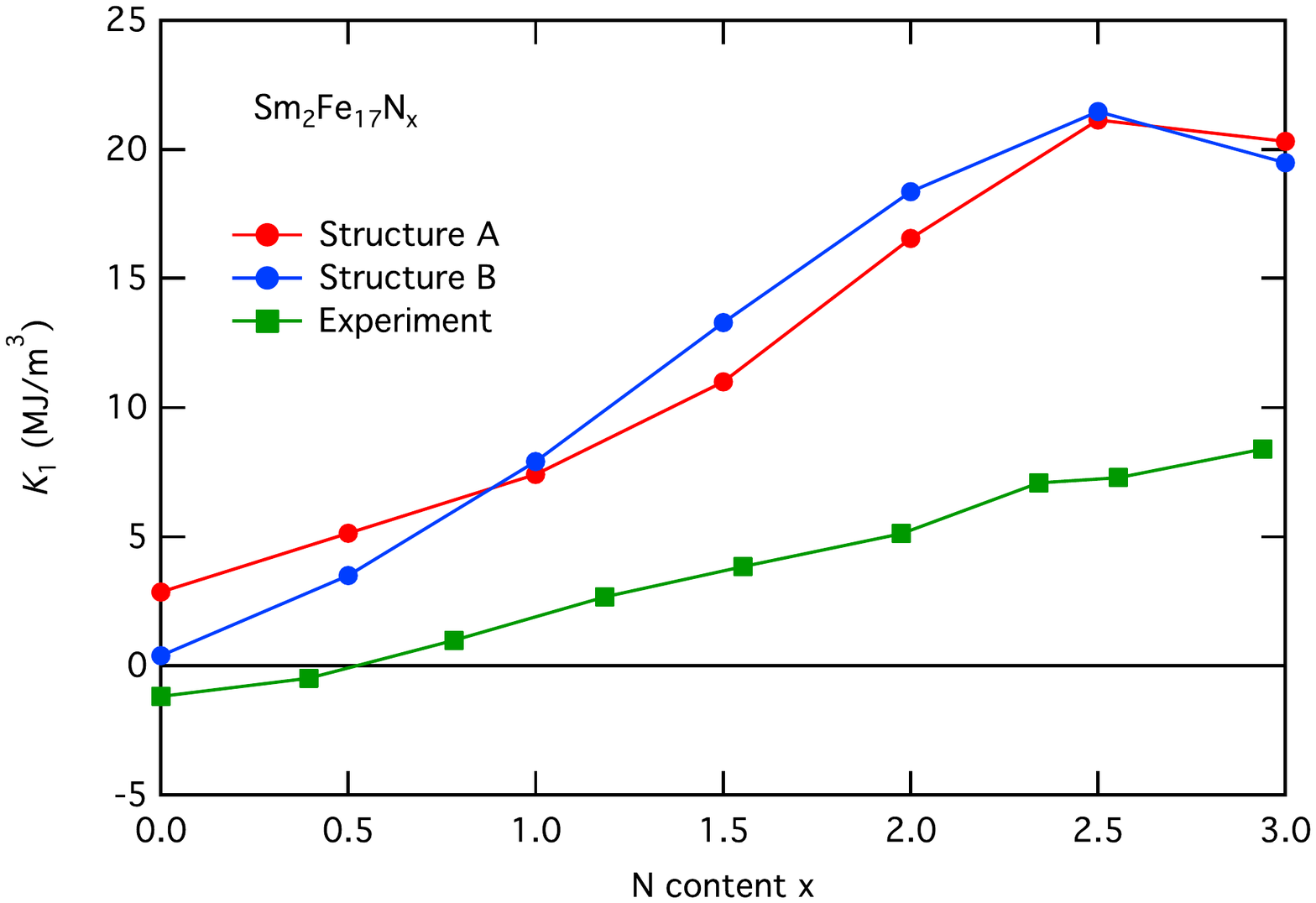}
\caption{(Color online) 
Magnetocrystalline anisotropy constant $K_1$ of Sm$_2$Fe$_{17}$N$_x$ as functions of N concentration $x$. The red and green circles show the results calculated for structures A and B, respectively.
}
\label{fig:smfen_k1}
\end{center}
\end{figure}

Figure~\ref{fig:smfen_k1} shows the calculated magnetocrystalline anisot\-ropy constant $K_1$ of Sm$_2$Fe$_{17}$N$_x$. Here, $K_1$ was evaluated from the total energy of the system as a function of the direction of the magnetization or, conversely, the direction of the crystal axes. The anisotropy energy is fitted to Eq.~(\ref{eq:MCA}).

The calculation was performed using the ASA, i.e., the crystal is filled with atomic spheres, the sum of whose volumes is the crystal volume, centered on each atomic site. Since the potential inside each atomic sphere is assumed to be spherically symmetric, electrons do not feel any anisotropic electrostatic field, namely, no crystal field effect arises. Therefore, in this calculation, all the magnetocrystalline anisotropy stems from the band structure that reflects the effects of SOC. This band structure effect, which is also understood as the effect of the hybridization with ligands, is distinguished from the crystal field effect. The former is usually more important than the latter for transition-metal ions but this is not necessarily the case for rare earths. In the present system, although it is a matter of course that the anisotropic electrostatic field also could be an important source of the magnetocrystalline anisotropy, the band structure effect makes a significant contribution. The overall trends of the behavior of magnetocrystalline anisotropy are reasonable. In particular, the behavior of $K_1$, starting from nearly zero at $x=0$, and increasing with increasing $x$, is well reproduced.  There is a considerable discrepancy in the absolute values of $K_1$ between the calculation and experiment. It is noted, however, that the calculated $K_1$ easily varies by $\pm 50$\% depending on the calculation details. For example, if the SIC procedure proposed by Perdew and Zunger \cite{perdew1981} is adopted instead of that by Filippetti and Spaldin \cite{filippetti2003}, which was used here, the value of $K_1$ becomes 10.2 MJ/m$^3$, which is considerably smaller than the present result of 20.3 MJ/m$^3$.

\begin{figure}[htp]
\vrule height 16pt width 0pt
\begin{center}
\includegraphics[width=9cm]{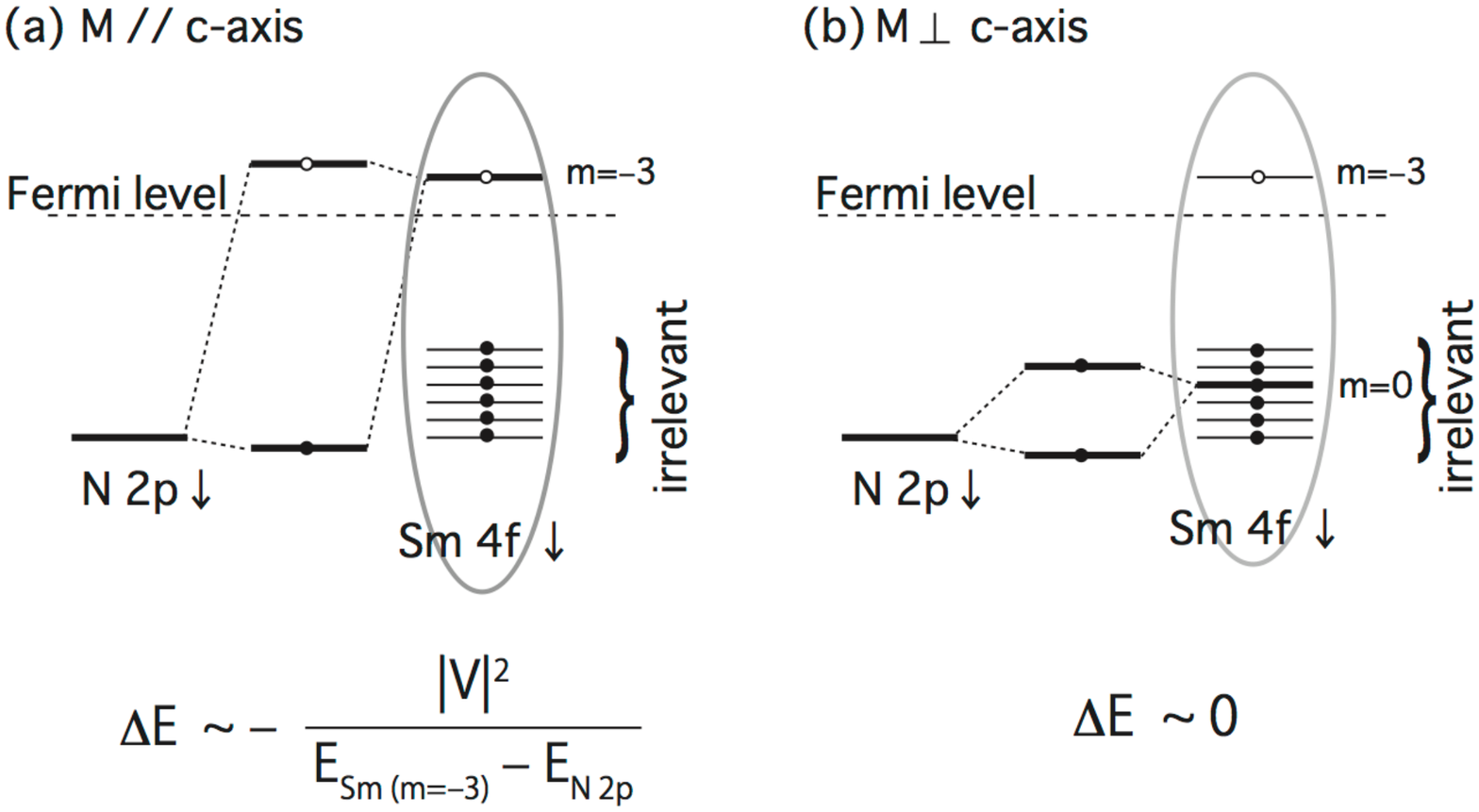}
\caption{
Energy diagram of N 2p and Sm 4f states and their hybridizations for magnetization (a) along $c$-axis and (b) perpendicular  to c-axis. Since the Sm f states for $-2\le m \le 3$ are fully occupied, the hybridization of these states with N 2p states does not contribute to the energy gain due to hybridization. Thus, only the $m=-3$ state contributes to the energy gain, which is the origin of the magnetic anisotropy in this case.
}
\label{fig:2p4f}
\end{center}
\end{figure}

The mechanism by which uniaxial anisotropy occurs is schematically shown in Fig. \ref{fig:2p4f}.  Without N atoms, the hybridization of Sm-f states with surrounding atoms is rather small and the f states keep a feature of narrow atomic-like state irrespective of the relative angle between the magnetization and crystal axes. In this situation, the rotation of crystal axes has little effect on the Sm-f states and causes no significant magnetic anisotropy. When N atoms are introduced, the hybridization between Sm-4f and N-2p states occurs. When the magnetization lies along the $c$-axis, the strongest hybridization occurs between N-2p and Sm-4f with magnetic quantum number $m=\pm 3$ states. On the other hand, for the in-plane magnetization, the strongest hybridization is between the Sm $m=0$ state and N-2p states. Comparing these two cases, we may say that an energy gain is expected only when the hybridization occurs between $m=-3$ and the N-2p states. This is because the SOC pushes up only the $m=-3$ state above the Fermi level and thus causes the energy gain due to the lowering of the occupied state energy levels. The importance of hybridization is also proven by the fact that, if the open-core scheme for Sm-f states be adopted, where no hybridization occurs between Sm-f and N-2p states, $K_1$  takes a small negative value.  The mechanism of such energy gains is the same as the superexchange working between two local magnetic moments: the virtual process to unoccupied states plays a role. The energy gain, and hence the magnetic anisotropy energy per N atom, due to hybridization is thus given by
\begin{equation}
\Delta E \sim  -\frac{|V|^2}{E_{\rm Sm(m=-3)} - E_{\rm N 2p}},
\end{equation}
where $V$ is the hybridization energy between the Sm-4f state with $m=-3$ of energy $E_{\rm Sm(m=-3)}$ and N-2p states of energy $E_{\rm N 2p}$. A similar  effect  caused by the hybridization between the unoccupied N-2p states and  occupied Sm-4f states also exists, and it actually counteracts the above mechanism, i.e., less hybridization for the case of magnetization along the $c$-axis. However, this would not affect the magnetic anisotropy significantly because the unoccupied N-2p states, which are the antibonding states formed between the N-2p and Fe-3d states, are orthogonal to the N-2p states that hybridize with Sm-4f states, and contribute little to this mechanism.

\section{Finite-Temperature Magnetism}
\label{sec:finiteT}
\subsection{Local moment disorder method}
There are no general methods so far to treat the finite-temperature magnetism of metallic systems from first principles. However, several schemes that can potentially incorporate finite-temperature magnetism into a first-principles approach have been proposed and even applied to permanent magnet materials. One of them, which represents the most recent developments, may be schemes using dynamical mean field theory (DMFT) combined with first-principles calculation\cite{Kotliar2006, Graanas2012, Lichtenstein2001, Zhu2014, Delange2017}. In this method the effects of electron correlation are treated locally more or less (depending on the solver) exactly in the framework of the local model Hamiltonian. The band structure is fully taken into account using the framework of first-principles calculation. 

Another conceivable way may be to apply the spin-fluctuation theories \cite{moriya1985} developed for the tight-binding model (or Hubbard model) to the Kohn--Sham equations. In the framework of the tight-binding model, a standard scheme to deal with the finite-temperature magnetism of itinerant electron systems is based on the functional integral method. This approach was first applied to the ferromagnetism of narrow d-bands  by Wang and co-workers \cite{wang1969, evenson1970}, and by Cyrott and co-workers \cite{cyrot1970, lacour-gayet1974} In this approach the Stratonovich--Hubbard transformation\cite{stratonovich1957, hubbard1959}, which maps the problem of an interacting electron system to that of a non-interacting system with fluctuating auxiliary fields that are to be integrated out in the sense of a functional integral, is used to calculate the grand potential\cite{hubbard1959}. 

Although Moriya et al. \cite{moriya1985} went a little further in the framework of the functional integral method for general discussion, most other works used the static, single-site, and saddle-point approximations in performing the procedure implied by this method. In particular, Hubbard \cite{hubbard1979a,hubbard1979b,hubbard1981} and Hasegawa \cite{hasegawa1979,hasegawa1980} independently developed the theory of ferromagnetism for Fe, Co, and Ni within the above approximations. Once these approximations are exploited, the procedure is reduced to the calculation of the electronic structure of random substitutional alloys. For this reason, such approaches are also called {\it alloy analogy}.

The so-called local moment disorder (LMD) method (often called the disordered local moment (DLM) method) is a typical scheme using the alloy analogy. The method is viewed as, but with a slight nuance, being based on the functional integral method. Since the method is equally  applied to the ground state, the use of the LMD method is not restricted to the study of the finite-temperature properties: the method was used by Jo\cite{jo1980, jo1981} to describe the quantum critical point of magnetic alloys in the tight-binding model and later used in the framework of KKR-CPA-LDA to discuss similar problems by Akai and co-workers \cite{akai1993}. The LMD approach combined with KKR-CPA-LDA was also applied to discuss the magnetism above $T_{\rm C}$ of Fe and Co by Oguchi et al.\cite{oguchi1983} and by Pindor et al.\cite{pindor1983}. The approach was further developed by Gyorffy et al.\cite{gyorffy1985} and Staunton and co-workers \cite{staunton1992}. The major difference between the calculations based on the LMD combined with KKR-CPA-LDA and the classical calculations in the tight-binding model is that the former is based on DFT, and hence, all the dynamical effects are assumed to be incorporated in that framework.

In the prototype LMD scheme, two local magnetic states, one aligned parallel to the magnetization, the other antiparallel, are considered and the system is supposed to be a random alloy composed of atoms of these two distinct local magnetic states. It may be said that it simulates the paramagnetic state above $T_{\rm C}$. The energy difference between the ferromagnetic and random alloy (LMD) states then gives an estimate of $T_{\rm C}$. If one assumes that the system is described by a Heisenberg model, $T_{\rm C}$ is given by $2/3$ of the energy difference per number of magnetic ions. Thus, calculated $T_{\rm C}$'s usually show reasonable correspondence with ones obtained by the scheme using $J_{ij}$ mentioned earlier in this section. The above alloy-analogy-type scheme is suitable above $T_{\rm C}$ where the rotational symmetry in the spin space is preserved, and hence the Ising-like treatment becomes exact in the single-site treatment. This scheme may still be feasible even below $T_{\rm C}$, where the local rotational symmetry in the spin space breaks, for the calculations of some insensitive quantities such as magnetization. However, this certainly is not true for the magnetic anisotropy, where the vector nature of spins is essential. In such cases, the directional distribution of spin in the whole solid angle has to be considered; each angle corresponds to each constituent atom of the {\it alloy}. 

One of the attempts that are of great relevance to the study of permanent magnet materials is the studies by Staunton and co-workers \cite{staunton2004, staunton2006} An important feature of their approach in the present context is that they include the SOC in the framework, which renders the calculation of the temperature dependence of the magnetic anisotropy tractable. The information obtained from such calculations can be utilized for other completely different approaches,  which will be explained in the following subsection.

\subsection{Spin-model analysis}
Another approach to finite-temperature magnetism is analysis using a spin model. An effective spin Hamiltonian of a rare-earth magnet compound is expressed as follows: 

\begin{eqnarray}
{\cal H} &  = & {\cal H}_{\rm T} + {\cal H}_{\rm R} + {\cal H}_{\rm RT}  + {\cal H}_{\rm ext} \;,
\label{eq:spinH}\\
{\cal H}_{\rm T} &  = &
  -2 \sum_{\left<i,j\right>\in {\rm T}} J_{ij}^{\rm TT} {\mathbf S}_{i}\cdot{\mathbf S}_{j}
-\sum_{i\in {\rm T}}D_{i}^{\rm T}\left(S_{i}^{z}\right)^2 \;, 
\label{eq:H_T}\\
{\cal H}_{\rm R} &  = & 
\sum_{i\in {\rm R}} \sum_{l} \tilde{\theta}_l^{J_i}A_{l,i}^{m_l} \langle r^l \rangle_i \hat{O}_{l.i}^{m_l} \;,
\label{eq:H_R}\\
{\cal H}_{\rm RT} &  = & 
 2 \sum_{\left<i,j\right>,i\in{\rm R},j\in {\rm T}}J_{ij}^{\rm RT}
(g_{J}-1) {\mathbf J}_{i}\cdot{\mathbf S}_{j} \;,
\label{eq:H_RT}\\
{\cal H}_{\rm ext} &  = & -\mu_0 \sum_i \bf{m}_i \cdot H_{\rm ext} \;.
\end{eqnarray}
Here, ${\cal H}_{\rm T}$ is the Hamiltonian in the $T$ sublattices. The first term in Eq. (\ref{eq:H_T}) is the magnetic exchange coupling between the $i$th and $j$th sites, and the second term is the single-ion anisotropy. ${\cal H}_{\rm R}$ represents the single-ion anisotropy at the $R$ site, where $\tilde{\theta}_l^{J_i}$ is the Stevens factor, $A_{l,i}^{m_l}$ is the crystal-field parameter, and $\hat{O}_{l.i}$ is the Stevens operator equivalent. 
${\cal H}_{\rm RT}$ is the exchange coupling between the $R$ site and $T$ site, where ${\bf J}_i$ is the total magnetic momentum at the $R$ site, and ${\bf S}_j$ is the spin momentum at the $T$ site. Finally, ${\cal H}_{\rm ext}$ expresses coupling between the external magnetic field ${\bf H}_{\rm ext}$ and the magnetic moment at the $i$th site, ${\bf m}_i$. Although this form of the Hamiltonian has been known for a long time, it is only recently that a quantitative calculation has been carried out.  Matsumoto et al. have evaluated the parameters in Eqs. (\ref{eq:spinH})-(\ref{eq:H_RT}) for NdFe$_{12}$N by first-principles calculation, and solved the derived spin Hamiltonian by the classical Monte Carlo method \cite{matsumoto2016}. They  found that the anisotropy field at high temperatures is sensitive to the magnetic exchange coupling between $R$ and $T$, namely $J_{ij}^{\rm RT}$ in Eq. (\ref{eq:H_RT}). 

\begin{figure}[htp]
\includegraphics[width=8.5cm]{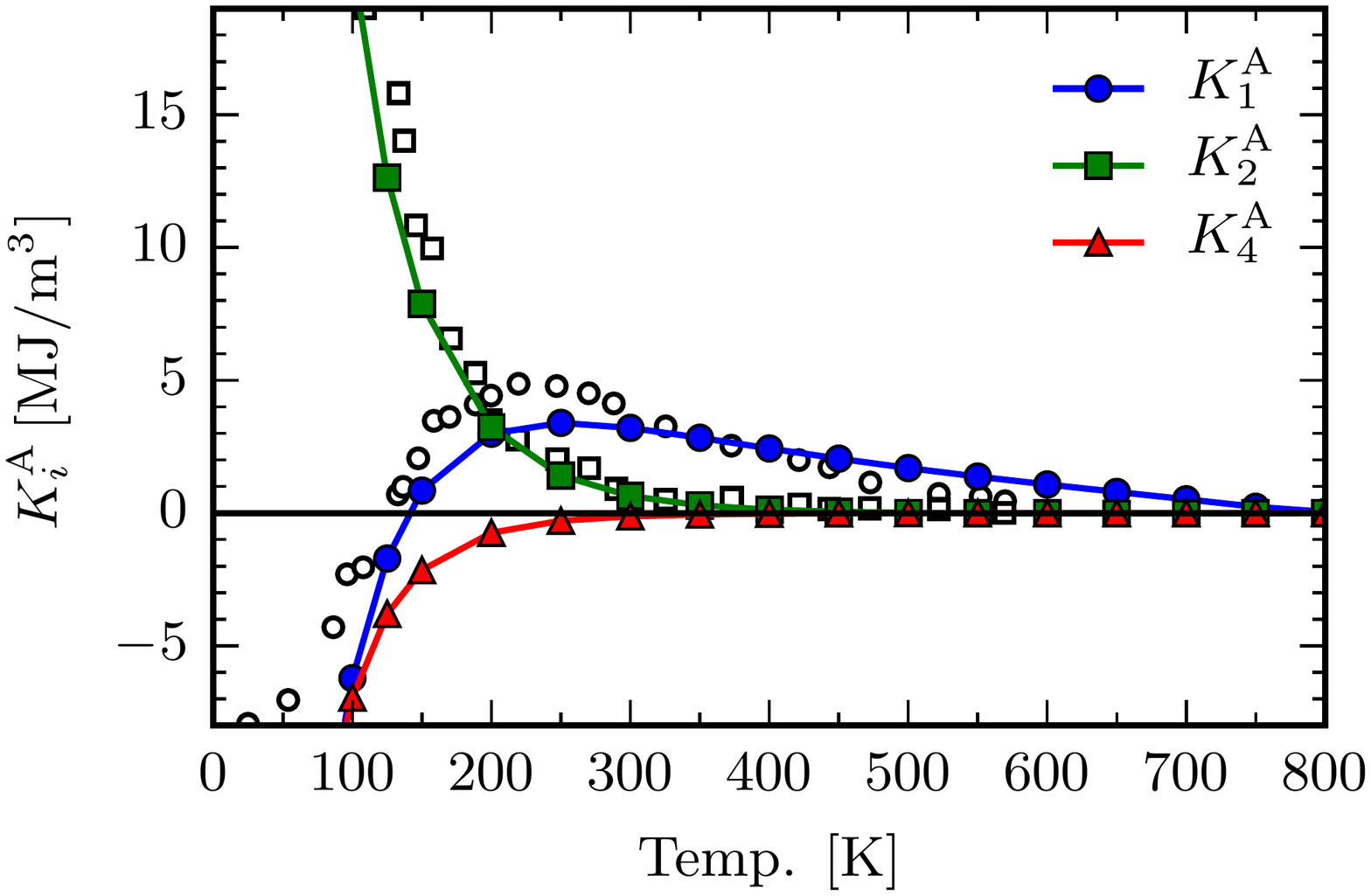}
\includegraphics[width=8.5cm]{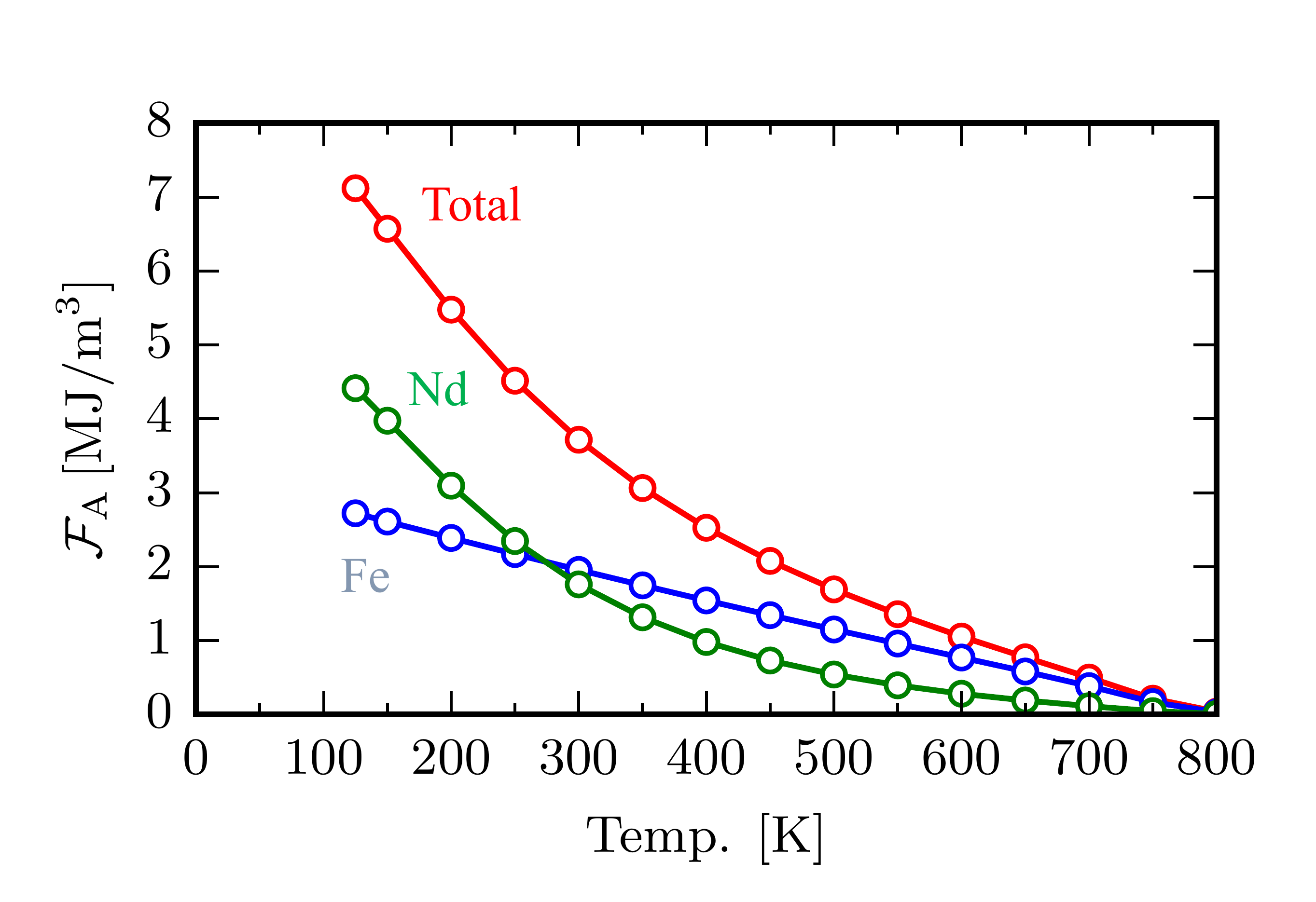}
\caption{(Color online) 
(a) Anisotropy constants $K_i^{\rm A}$ ($i$=1,2,3) of Nd$_2$Fe$_{14}$B obtained by constrained Monte Carlo simulations  for the classical spin Hamiltonian. Experimental values for $K_1^{\rm A}$ and $K_2^{\rm A}$ are also plotted for comparison. (b) Temperature dependence of the magnetic anisotropy energy. 
The figures are taken from ref.~[\citen{toga2016}].
Copyright 2016 by the American Physical Society.
}
\label{fig:toga}
\end{figure}

A similar simulation was carried out for Nd$_2$Fe$_{14}$B by Toga et al. \cite{toga2016}. They computed effective parameters from first principles except for $A_{l}^{m}$.  For the crystal-field parameters, experimentally deduced values for $l$=2,4,6 and $m$=0 \cite{yamada1988} were used. The calculated magnetization obtained by Monte Carlo simulation  successfully reproduced the spin reorientation transition at $\sim$140 K. The Curie temperature was calculated to be 754 K, which is in reasonable agreement with the experimental value of 585 K. The magnetocrystalline anisotropy energy can be  computed using the constrained Monte Carlo method \cite{asselin2010}. In this method, the direction of the total magnetization is fixed to a given angle $\theta$. The direction of the spin magnetic moment at each site is changed under this constraint, and the thermal average is taken by Monte Carlo simulation. The free energy $\cal{F}(\theta)$ is then obtained from 

\begin{equation}
{\cal F}(\theta) = \int^{\theta} d \theta' [ {\bf n}(\theta') \times {\cal{\bf T}} (\theta') ] \cdot \frac{\partial {\bf n}(\theta)}{\partial \theta} \;,
\end{equation}
where $\cal{\bf T}(\theta)$ is the magnetic torque and ${\bf n}(\theta)$ is the unit vector in the direction of the total magnetization. Figure~\ref{fig:toga}(a) shows the magnetic anisotropy constants obtained by fitting to the following equation:

\begin{equation}
{\cal F}(\theta, T) = K_1^{\rm A}(T) \sin^2\theta + K_2^{\rm A}(T) \sin^4\theta + K_3^{\rm A}(T) \sin^6\theta + {\rm const.} \;,
\end{equation}
where $T$ is the temperature. We see that both $K_1^{\rm A}(T)$ and $K_2^{\rm A}(T)$ are in good agreement with the experiment \cite{Yamada1986} except for $T <$ 100 K, where the quantum effect would become significant. Figure \ref{fig:toga}(b) shows the magnetocrystalline anisotropy energy ${\cal F}_{\rm A}$ as a function of temperature. Contributions from Nd and Fe sites are also plotted, which are evaluated by hypothetically putting $D^{\rm T}$=0 and $A_l^m$=0, respectively. The anisotropy energy becomes weaker as the temperature increases. At low temperatures, the magnetic anisotropy originating from Nd sites is stronger than that from Fe sites. This is naturally understood because strong magnetocrystalline anisotropy originates from single-ion anisotropy at $R$ sites. As the temperature is raised, however, the Nd contribution decays quickly, whereas the Fe contribution decreases gradually with linear dependence against temperature. The quick decay of the Nd contribution was explained by Sasaki et al. in the molecular-field approximation as follows \cite{sasaki2015}. The exchange field acting on $R$ from $T$ becomes weaker with increasing temperature because of thermal fluctuation. Then, the energy splitting between the $m$ states in the $J$ multiplet becomes smaller. Excited $m$ states are easily occupied by thermal excitation, and the 4f electron distribution approaches a spherical distribution. 
As a consequence, the crystal-field effect becomes ineffective, which results in a decrease in the magnetic anisotropy at $R$. This implies that the magnetic exchange field is more important than the single-ion anisotropy for magnetic anisotropy at high temperatures. 

\section{$R$Fe$_{12}$-Type Compounds}
\label{sec:RFe12}

$R$Fe$_{12}$-type compounds with the ThMn$_{12}$ structure have been investigated actively in the past few years. This class of compounds was studied as possible strong magnet compounds in the late 80's when iron-rich phases were synthesized \cite{buschow1987a,buschow1988a,Buschow1991}. Among them, SmFe$_{11}$Ti was developed by Ohashi et al. \cite{ohashi1987}. Subsequently, it was found that interstitial nitrogenation improves magnetic properties: the magnetization is enhanced and the Curie temperature rises by 100--200 K. Magnetocrystalline anisotropy is also affected significantly.  Yang and co-workers found that NdFe$_{11}$TiN$_{\delta}$ is a good magnet compound having reasonably high saturation magnetization \cite{yang1991a,yang1991b}. However, the magnetization is smaller than that of Nd$_2$Fe$_{14}$B, which had already been developed. Hence, $R$Fe$_{12}$-type compounds have not been studied extensively for two decades. 

\begin{figure*}[htp]
\includegraphics[width=18cm]{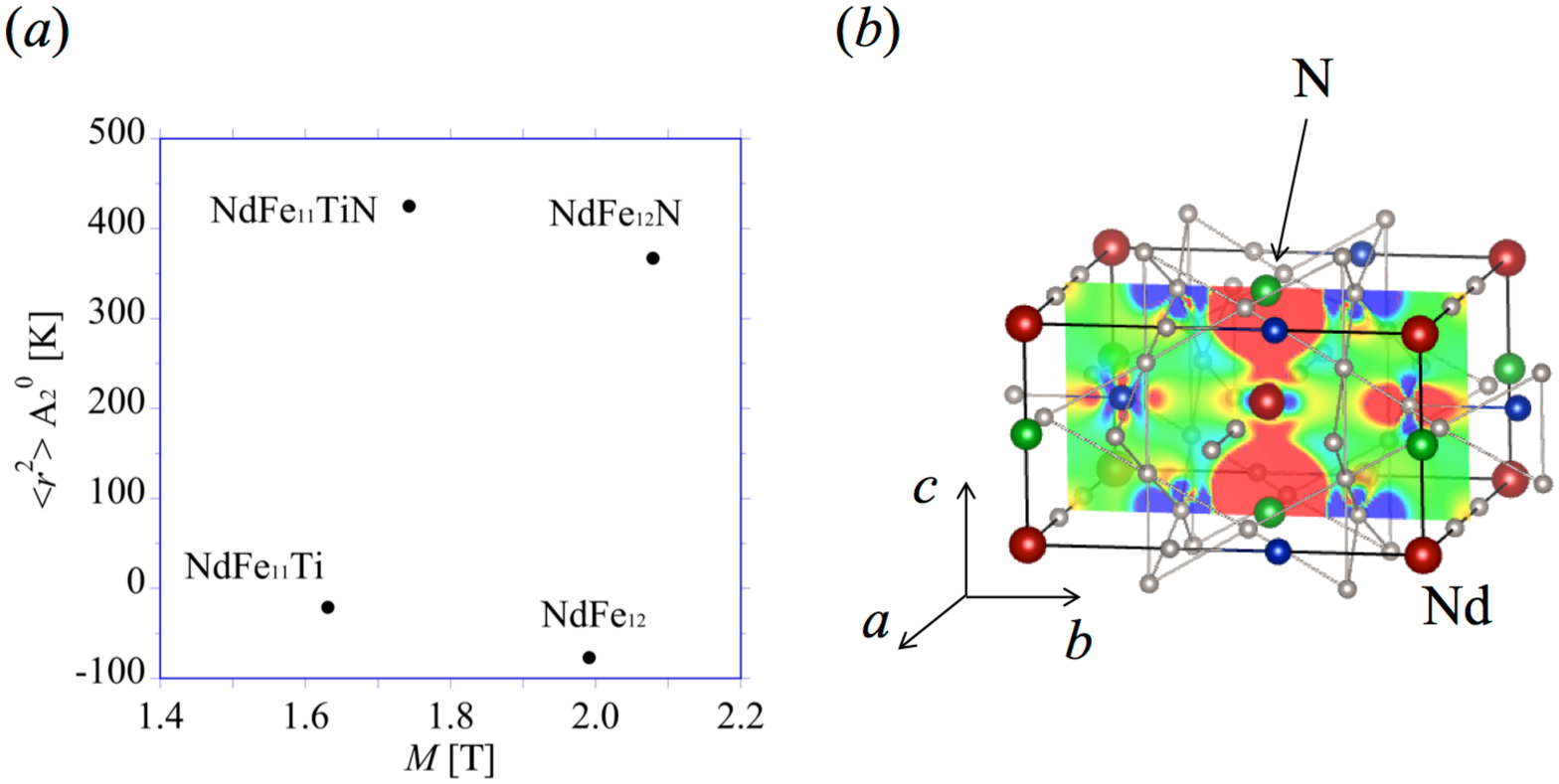}
\caption{(Color online) 
(a) Magnetization and second-order crystal-field parameter of NdFe$_{11}$Ti, NdFe$_{12}$, NdFe$_{11}$TiN, and NdFe$_{12}$N. (b) Difference in the electron density between NdFe$_{11}$TiN and NdFe$_{11}$Ti. In the latter case, nitrogen is removed from the former by a fixing structure \cite{miyake2014,harashima2015b}. 
The electron density increases (decreases) by $>$ 0.001 /(Bohr)$^3$ in the red (blue) region.
}
\label{fig:miyake}
\end{figure*}

The $R$Fe$_{12}$-type compounds contain a high Fe content. This is advantageous for achieving large saturation magnetization. Nevertheless, NdFe$_{11}$TiN has lower magnetization than Nd$_2$Fe$_{14}$B because of the presence of Ti, which substitutes for one of Fe sites. Miyake and co-workers studied NdFe$_{11}$TiN and NdFe$_{12}$N by first-principles calculation \cite{miyake2014, harashima2015b}. They found that NdFe$_{11}$TiN has substantially smaller magnetization than NdFe$_{12}$N because (1) the spin is negatively polarized at the Ti site, and (2) the magnetic moments at Fe sites in the vicinity of Ti are suppressed on average. As a result, the reduction of the magnetization by Ti substitution is more significant than the naive expectation from the change in the iron concentration. On the other hand, the $A_{2}^0$ parameter in NdFe$_{12}$N is comparable to that of NdFe$_{11}$TiN, suggesting that NdFe$_{12}$N has reasonably large magnetocrystalline anisotropy [Fig.\ref{fig:miyake}]. In both NdFe$_{11}$TiN and NdFe$_{12}$N, interstitial nitrogenation enhances the $A_{2}^0$ parameter drastically. This is because a weak chemical bond is formed between Nd and N, and the electron density increases between them [Fig.\ref{fig:miyake}(b)]. It pushes away the Nd-4f electrons in the perpendicular direction, which induces uniaxial magnetocrystalline anisotropy. Subsequently, Hirayama and co-workers synthesized NdFe$_{12}$N on a MgO substrate with a W underlayer, and reported that the compound has larger saturation magnetization and anisotropy field than Nd$_2$Fe$_{14}$B \cite{hirayama2015}. 

\begin{figure}[htp]
\includegraphics[width=9cm]{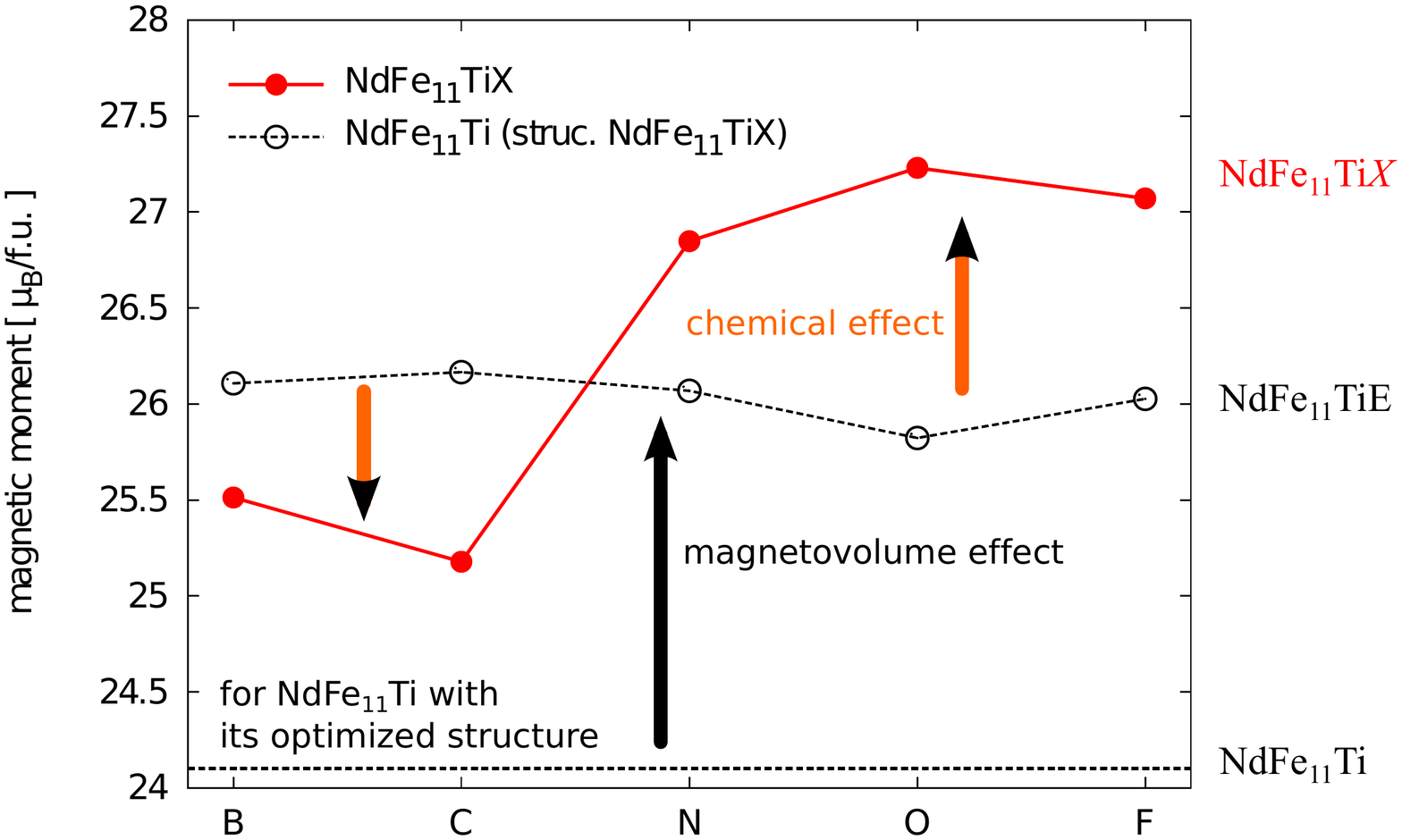}
\caption{(Color online) 
Calculated spin magnetic moment and magnetization (per volume) of NdFe$_{11}$Ti and NdFe$_{11}$Ti$X$ for $X$=B, C, N, O, and F. NdFe$_{11}$Ti$E$ is the result for NdFe$_{11}$Ti in which $X$ is removed from NdFe$_{11}$Ti$X$ by a fixing structure. The figure is taken from ref.~[\citen{harashima2015a}].
Copyright 2015 by the American Physical Society.
}
\label{fig:harashima}
\end{figure}

Interstitial light elements have a strong influence on the magnetism of rare-earth magnet compounds. Kanamori discussed the role of B in Nd$_2$Fe$_{14}$B \cite{kanamori1990,kanamori2006}. When B is added to iron compounds, the B-2p state hybridizes with the Fe-3d states. Since the B-2p energy level is located higher than the Fe-3d level, the antibonding state, having strong B-2p character, appears above the Fermi level. Then, the Fe-3d state is pushed down by p-d hybridization. This suppresses the spin magnetic moment of Fe sites neighboring B. This is called {\it cobaltization}. Meanwhile, the 3d orbital at cobaltized Fe sites hybridizes with 3d orbitals at surrounding Fe sites. Then, the spin magnetic moment is enhanced at the latter sites. These chemical effects can have a sizable effect on the total magnetization of the compound. Harashima et al. studied these effects in the hypothetical compound NdFe$_{11}$TiB \cite{harashima2015a}. They confirmed that the change in the spin magnetic moment at each Fe site is explained by the cobaltization mechanism. They also found that the net change in the total spin magnetic moment is negative, namely the chemical effect induced by B reduces the total magnetic moment. As a matter of fact, the total magnetic moment of NdFe$_{11}$TiB is larger than that of NdFe$_{11}$Ti, but this is attributed to magnetovolume effect [Fig.\ref{fig:harashima}]. As the light element $X$ is changed from B to C or N, the magnetic moment shows a  jump between $X$=C and $X$=N. A similar result has been reported in related systems \cite{akai1995,asano1997}. This $X$ dependence originates from a chemical effect. The antibonding state between the $X$-2p and Fe-3d states is downshifted as the atomic number of $X$ increases. Eventually, the hybridized state crosses the Fermi level. It  is partially occupied in the majority-spin channel for $X$=N, leading to enhancement of the magnetic moment. (The symmetry of the hybridized state changes from antibonding to bonding character as the state crosses the Fermi level.)

An important issue of $R$Fe$_{12}$-type compounds is how to stabilize the bulk phase. This is achieved by substituting  part of the Fe sites with another element $M$ such as Ti, V, Cr, Mn, Mo, W, Al, or Si. However, these stabilizing elements decrease the magnetic moment. The search for a stabilizing element that does not lead to significant magnetization reduction is a  hot topic \cite{harashima2016}. 

\section{Concluding remarks}
\label{sec:summary}
Although research on rare-earth magnets has a long history, the quantitative understanding is still insufficient. A theoretical framework of first-principles calculation is under development, mainly because of the difficulty in treating the 4f electrons of rare-earth elements. Accurate description of magnetism is an important issue yet to be resolved. Another challenge is applications to grain boundaries. Real permanent magnets contain additive elements, impurities, defects, and various subphases. Of particular interest are the interfaces between the main phase and grain boundary phases, which are believed to play a crucial role in coercivity. The recent development of supercomputers enables us to directly compute the interfaces by first-principles calculation. Exploration of a new magnet is also a major challenge. The neodymium-based magnet has been the strongest permanent magnet for the last thirty years. NdFe$_{12}$N has superior intrinsic magnetic properties; however, its thermodynamic instability prevents industrial application. Exploration of a wide range of compounds is anticipated. The recent development of materials-informatics  may help us discover new magnet compounds efficiently. 
\\

\begin{acknowledgment}
We acknowledge collaboration and fruitful discussions with Kiyoyuki Terakura, Yosuke Harashima, Hiori Kino, Shoji Ishibashi, Taro Fukazawa, Shotaro Doi, Munehisa Matsumoto, Yuta Toga, Seiji Miyashita, Akimasa Sakuma, Masako Ogura and Satoshi Hirosawa. 
This work was partly supported by the Elements Strategy Initiative Project under the auspices of MEXT; by MEXT as a social and scientific priority issue (Creation of New Functional Devices and High-Performance Materials to Support Next-Generation Industries; CDMSI) to be tackled by using a post-K computer; 
by Grants-in-Aid for Scientific Research, MEXT 26400330 and 17K05556; and also 
by the ``Materials Research by Information Integration'' Initiative (MI$^2$I) project of the Support Program for Starting Up Innovation Hub from the Japan Science and Technology Agency (JST). 
\end{acknowledgment}

\bibliographystyle{jpsj}
\bibliography{./ref}

\end{document}